\documentclass[prd,amsmath,twocolumn,floatfix,showpacs,preprintnumbers]{revtex4}
\usepackage{epsfig}
\usepackage{graphicx}

\newcommand{\rf}[4]{{#1} {\bf #2}, #3 (#4)}

\newcommand{\physl}{Phys.\ Lett.}

\newcommand{\np}{Nucl.\ Phys.}
\newcommand{\npps}[3]{Nucl.\ Phys. {\bf B} (Proc.\ Suppl.) {\bf #1},
        #2 (#3)}
\newcommand{\cm}{Commun.\ Math.\ Phys.}
\newcommand{\etal}{\emph{et al.}}
                                                                                                                                               
\newcommand{\del}{\partial}
\newcommand{\lapI}{$\partial^2$(I) }
\newcommand{\lapII}{$\partial^2$(II) }
\newcommand{\lapIII}{$\partial^2$(III) }

\newcommand{\adj}{\dagger}
\newcommand{\tr}{\text{Tr}}

\newcommand{\oa}[1]{\ensuremath{{\cal O}(a^{#1})}}

\newcommand{\eref}[1]{Eq.~(\ref{#1})}

\newcommand{\be}{\begin{equation}}
\newcommand{\ee}{\end{equation}}
\newcommand{\slh}{\!\!\!\slash}

\begin{document}

\preprint{ADP-04-023/T605}

\title{
Quark propagator in Landau and Laplacian gauges with overlap fermions
}
\author{J.\ B.\ Zhang$^{a}$, Patrick O.\ Bowman$^{a,b}$, Ryan J.\ Coad$^{a}$, Urs M.\ Heller$^{c}$,
Derek B.\ Leinweber$^{a}$, and Anthony G.\ Williams$^{a}$}
\affiliation{
$^{a}${CSSM Lattice Collaboration, \\
Special Research Center for the Subatomic Structure of
Matter (CSSM) and Department of Physics,
University of Adelaide 5005, Australia} \\
$^{b}${Nuclear Theory Center, Indiana University, Bloomington IN 47405,
USA} \\
$^{c}${American Physical Society, One
Research Road, Box 9000, Ridge, NY 11961-9000, USA}
}
                                                                                                                                     
\date{\today}
                                                                                                                                     
                                                                                                                                     
\begin{abstract}
The properties of the momentum space quark propagator in Landau gauge and Gribov copy
free Laplacian gauge are studied for the overlap quark action in quenched lattice QCD.
Numerical calculations are done on two lattices with different  lattice
spacing $a$ and the same physical volume. We have calculated the
nonperturbative wave function renormalization function $Z(q)$
and the nonperturbative mass function $M(p)$ for a variety of bare quark masses
and perform a simple linear extrapolation to the chiral limit. We focus on the comparison
of the behavior of $Z(q)$ and $M(p)$ in the chiral limit in the two gauge fixing schemes as well as 
the behavior on two lattices with different  lattice spacing $a$. We find that the
mass functions $M(p)$ are very similar for the two gauges while the wave-function renormalization function $Z(q)$
is more strongly infrared suppressed in the Laplacian gauge than in the Landau gauge on the 
finer lattice. For Laplacian gauge, it seems that the finite $a$ error is large on the coarse
lattice which has a lattice spacing $a$ of about 0.124 fm.
\end{abstract}

\pacs{ 
12.38.Gc,  
11.15.Ha,  
12.38.Aw,  
14.65.-q   
}

\maketitle

\section{Introduction}
Quantum Chromodynamics (QCD) is widely accepted as the correct theory of the 
strong interaction and 
the quark propagator is one of its fundamental quantities. By studying 
the momentum-dependent quark mass function in the infrared region we can gain 
valuable insight into the mechanism of dynamical chiral symmetry breaking and 
the associated dynamical generation of mass. At high momenta, one can use
the quark propagator to extract the running quark mass~\cite{Bowman:2004xi}.  

Lattice QCD provides
a way to study the quark propagator nonperturbatively.   
There have been several  lattice studies of the momentum space quark 
propagator~\cite{soni1,soni2,jon1,jon2,Bow02a,Bow02b,blum01,overlgp,overlgp2,qpscaling} 
using different fermion actions.  
The usual gauge
for these studies has been Landau gauge, because it is a (lattice) Lorentz
covariant gauge that is easy to implement on the lattice, and  
the results from the lattice Landau gauge can be easily compared to studies 
that use different methods.  Finite volume effects
and discretization errors have been extensively explored in lattice Landau
gauge~\cite{overlgp2,qpscaling}. 
Unfortunately, lattice Landau gauge suffers from the well-known problem of
Gribov copies.  Although the ambiguity originally noticed by
Gribov~\cite{Gri78} is not present on the lattice, since in practice one never 
samples from the same gauge orbit twice, the
maximization procedure used for gauge fixing does not uniquely fix the gauge.
In general, there are many local maxima for the algorithm to choose from,
each one corresponding to a Gribov copy, and no local algorithm can choose the
global maximum from among them.  While various remedies have
been proposed~\cite{Het98,evolutionary}, they are either unsatisfactory or
computationally very intensive.  For a recent discussion of the Gribov problem
in lattice gauge theory, see Ref.~\cite{Wil02}.

An alternative approach is to work with the so-called
Laplacian gauge~\cite{Vin92}.  This gauge is ``Landau like'' in the sense that 
it has
similar smoothness and Lorentz invariance properties~\cite{Vin95}, but it
involves a non-local gauge fixing procedure that avoids lattice Gribov copies.
The gluon propagator has already been studied in Laplacian gauge in
Refs.~\cite{Ale01,lapgp} and the improved staggered quark
propagator in Laplacian gauge in Ref.~\cite{Bow02a}. It has been 
shown~\cite{baal95} that Landau and Laplacian gauges become equivalent in the 
perturbative (high-momentum) regime and this has
been confirmed by numerical studies~\cite{Bow02a,Ale01,lapgp}.

In this paper we study the overlap quark propagator in the Laplacian
gauge and compare the results with the Landau gauge to explore the
effects of selecting a gauge condition free of Gribov copies.  Unlike
Asqtad fermions, the overlap formalism provides a fermion action which
is free of doublers and preserves an exact form of chiral symmetry on
the lattice.  The latter feature makes overlap fermions the action of
choice for studying dynamical chiral symmetry breaking near the chiral limit.

We also compare Laplacian gauge results on two lattices with the same
physical volume and different lattice spacings $a$ to explore the
finite $a$ error.  In this work, the \oa{2} mean-field improved gauge
action is used to generate the quenched gauge configurations.

\section{Gauge Fixing}
\label{sec:gauge}

We consider the quark propagator in Landau and Laplacian gauges.  Landau gauge
fixing is performed by enforcing the Lorentz gauge condition,
$\sum_\mu \del_\mu A_\mu(x) = 0$ on a configuration by configuration basis.
For the tadpole improved plaquette plus
rectangle (L\"{u}scher-Weisz~\cite{Lus85}) gauge action which we use in the current work, 
we use the \oa{2} improved gauge fixing scheme, this is achieved by maximizing the functional~\cite{bowman2},
\begin{equation}
{\cal F} = \frac{4}{3}{\cal F}_{1} - \frac{1}{12u_0}{\cal F}_{2}, 
\end{equation}
where ${\cal F}_{1}$ and  ${\cal F}_{2}$ are
$$
{\cal F}_1 = \frac{1}{2} \sum_{x,\mu} \tr \bigl\{ U_\mu(x)
        + U_\mu^\adj(x) \bigr\}
$$
and 
$$
{\cal F}_2 = \frac{1}{2} \sum_{x,\mu} \tr \bigl\{ U_\mu(x)U_\mu(x+\mu)
        + U_\mu^\adj(x+\mu) U_\mu^\adj(x) \bigr\}
$$
respectively, and $u_0$ is the usual plaquette measure of the mean link. 
In this case, a Fourier accelerated, steepest-descents
algorithm~\cite{Dav88} is used to find a local maximum. There are, in general, many local 
maxima and these are called lattice Gribov copies.  This ambiguity
in principle will remain a source of uncontrolled systematic error.

Laplacian gauge fixing is a nonlinear gauge fixing that respects rotational
invariance, has been seen to be smooth, yet is free of Gribov ambiguity.
It is also computationally cheaper then Landau gauge fixing.  There is, however, more
than one way of obtaining such a gauge fixing in SU(N) lattice gauge theory.
There are three implementations of Laplacian gauge fixing employed in the literature:
\begin{enumerate}
\item \lapI gauge (QR decomposition), used by Alexandrou
        \etal~\cite{Ale01}.
\item \lapII gauge, where the Laplacian gauge transformation is projected
        onto SU(3) by maximising its trace~\cite{lapgp}.
\item \lapIII gauge (Polar decomposition), the original prescription described
        in Ref.~\cite{Vin92} and tested in Ref.~\cite{Vin95}.
\end{enumerate}
All three versions reduce to the same gauge in SU(2).  For a more detailed
discussion, see Ref.~\cite{lapgp}. For SU(3) staggered quarks, the study in Ref.~\cite{Bow02a}
indicate that \lapI and \lapII gauge give very similar results, and \lapIII gauge
is very noisy. In this work we will only use the \lapII gauge.

\section{Quark Propagator on the Lattice}
\label{lattice}

In a covariant gauge in the continuum, the renormalized Euclidean
space quark propagator has the form
\begin{eqnarray}
S(\zeta^2;p)=\frac{1}{i {p \slh} A(\zeta^2;p^2)+B(\zeta^2;p^2)}
=\frac{Z(\zeta^2;p^2)}{i{p\slh}+M(p^2)}\, ,
\label{ren_prop}
\end{eqnarray}
where $\zeta$ is the renormalization point. The renormalization 
point boundary conditions are chosen to be
\begin{equation}
Z(\zeta^2;\zeta^2)\equiv 1 \qquad M(\zeta^2)\equiv m(\zeta^2) \, .
\end{equation}
where $m(\zeta^2)$ is the renormalized quark mass at the renormalization point.
The functions $A(\zeta^2;p^2)$ and $B(\zeta^2;p^2)$, or alternatively
$Z(\zeta^2;p^2)$ and $M(p^2)$,  contain all of the nonperturbative information
of the quark propagator.  Note that $M(p^2)$ is renormalization
point independent, 
all of the renormalization-point dependence is carried by
$Z(\zeta^2;p^2)$. 

When all interactions for the quarks are turned off, i.e., when the gluon
field vanishes (or the links are set to one), the quark propagator has its 
tree-level form
\begin{equation}
S^{(0)}(p)=\frac{1}{i{p\slh}+m^0} \, ,
\end{equation}
where $m^0$ is the bare quark mass.  When the interactions with the
gluon field are turned on we have
\begin{equation}
S^{(0)}(p) \to S^{\rm bare}(a;p) = Z_2(\zeta^2;a) S(\zeta^2;p) \, ,
\label{tree_bare_ren}
\end{equation}
where $a$ is the regularization parameter - in this case, the lattice spacing 
- and $Z_2(\zeta^2;a)$ is the quark wave-function renormalization constant
chosen so as to ensure $Z(\zeta^2;p^2)|_{p^2 = \zeta^2}=1$.  For simplicity of notation we 
suppress the $a$-dependence of the bare quantities.

On the lattice we expect the bare quark propagators, in momentum space,
to have a similar form as in the continuum, except
that the $O(4)$ invariance is replaced by a 4-dimensional hypercubic
symmetry on an isotropic lattice.
Hence, the inverse lattice bare quark propagator takes the general form
\begin{equation}
(S^{\rm bare})^{-1}(p)\equiv
{i\left(\sum_{\mu}C_{\mu}(p)\gamma_{\mu}\right)+B(p)}.
\label{invquargen}
\end{equation}
With the periodic boundary conditions in the spatial directions
and anti-periodic in the time direction,
the discrete lattice momenta will be 
\begin{eqnarray}
p_i=\frac{2\pi}{N_{i}a}\left(n_i-\frac{N_i}{2}\right),\hspace{0.2cm}{\rm{and}}\hspace{0.2cm}p_t
=\frac{2\pi}{N_{t}a}\left(N_t-\frac{1}{2}-\frac{N_t}{2}\right) \, ,
\label{dismomt}
\end{eqnarray}
where $n_i=1,..,N_i$ and $n_t=1,..,N_t$, $N_i$ and $N_t$ are the lattice extent in spatial and temporal direction
respectively.

The overlap fermion formalism~\cite{neuberger0,neuberger2}
realizes an exact chiral
symmetry on the lattice and is automatically ${\cal O}(a)$ improved.
The massive overlap operator can be written as~\cite{edwards2}
\begin{eqnarray}
D(\eta) = \frac{1}{2}\left[1+\eta+(1-\eta)\gamma_5 \epsilon(H_w) \right] \, ,
\label{D_mu_eqn}
\end{eqnarray}
where $H_w(x,y)=\gamma_5 D_w(x,y)$ is the Hermitian Wilson-Dirac
operator,  $\epsilon(H_w)$ = $H_w/\sqrt{H_w^2}$ is the matrix sign function,
and the dimensionless quark mass parameter $\eta$ is
\begin{equation}
\eta \equiv \frac{m^0}{2m_w} \, ,
\label{mu_defn}
\end{equation}
where $m^0$ is the bare quark mass and $m_w$ is the Wilson quark mass which,
in the free case, must be in the range $0 < m_w < 2$.  
The bare quark propagator in coordinate space is given by the equation
\begin{equation}
S^{\rm bare}(m^0)\equiv \tilde{D}_c^{-1}(\eta) \, ,
\label{overlap_propagator}
\end{equation}
where 
\begin{equation*}
\tilde{D}_c^{-1}(\eta) \equiv \frac{1}{2m_w} \tilde{D}^{-1}(\eta) \hspace{0.5cm}{\rm{and}}
\end{equation*}
\be
\tilde{D}^{-1}(\eta) \equiv \frac{1}{1-\eta}\left[{D}^{-1}(\eta)-1\right]
\, .
\label{D_mu}
\end{equation}
 
When all the interactions are turned off, the inverse bare lattice quark propagator becomes the 
tree-level version of \eref{invquargen}
\begin{equation}
(S^{(0)})^{-1}(p)\equiv
{i\left(\sum_{\mu}C_{\mu}^{(0)}(p)\gamma_{\mu}\right)+B^{(0)}(p)}\,.
\label{treeinvpro}
\end{equation}
We calculate $S^{(0)}(p)$ directly by setting the links to unity in the
coordinate space, doing the matrix inversion and then taking its Fourier transform.
It is then possible to identify the appropriate kinematic lattice
momentum $q$  directly from the definition
\begin{equation}
q_\mu\equiv C_{\mu}^{(0)}(p).
\label{latmomt}
\end{equation}
The form of $q_\mu(p_\mu)$ is shown and its analytic form given in 
Ref.~\cite{overlgp}.  Having identified the appropriate kinematical lattice 
momentum $q$, we can now define the bare lattice propagator as
\begin{equation}
S^{\rm bare}(p)
\equiv \frac{Z(p)}{i{q\slh}+M(p)}.
\end{equation}
This ensures that the free lattice propagator is identical to the free 
continuum propagator.  Due to asymptotic freedom the lattice propagator will
also take the continuum form at large momenta.  In the gauge sector, this
type of analysis dramatically improves the gluon 
propagator~\cite{Lei99,Bon00,Bon01}.

The two Lorentz invariants can then obtained by
\begin{gather}
Z^{-1}(p) = \frac{1}{12iq^2} \tr \{q\slh S^{-1}(p) \} \\
M(p) = \frac{Z(p)}{12} \tr \{ S^{-1}(p) \}.
\end{gather}
This means that $Z(p)$ is directly dependent on our choice of momentum, $q$, while
$M(p)$ is not.

\section{NUMERICAL RESULTS}
\label{numerical}

\subsection{Simulation parameters}
 In this paper we work on two lattices with different lattice spacing, $a$, and very 
similar physical volumes. The gauge configurations are created using a tadpole improved plaquette plus 
rectangle (L\"{u}scher-Weisz~\cite{Lus85}) gauge action through the 
pseudo-heat-bath algorithm.  For each lattice size,  
50 configurations are used.
Lattice parameters are summarized in Table~\ref{simultab}.  The lattice
spacing $a$ is determined from the static quark potential with a string
tension $\sqrt{\sigma}=440$~MeV~\cite{zanotti}.

\begin{table*}[ht]
\caption{\label{simultab}Lattice parameters.}
\begin{ruledtabular}
\begin{tabular}{cccccccc}
Action &Volume &$N_{\rm{Therm}}$ & $N_{\rm{Samp}}$ &$\beta$ &$a$ (fm) & $u_{0}$ & Physical Volume (fm$^4$)\\
\hline
Improved       & $16^3\times{32}$ & 5000 & 500 & 4.80 & 0.093  & 0.89650 & $1.5^3\times{3.00}$ \\
Improved       & $12^3\times{24}$ & 5000 & 500 & 4.60 & 0.124  & 0.88888 & $1.5^3\times{3.00}$ \\
\end{tabular}
\end{ruledtabular}
\end{table*}

Landau gauge fixing to the gauge configuration was done using a
Conjugate Gradient Fourier Acceleration~\cite{cm} algorithm with an accuracy
of $\theta\equiv\sum\left|\partial_{\mu}A_{\mu}(x)\right|^{2}<10^{-12}$.
The improved gauge-fixing scheme was used to minimize
gauge-fixing discretization errors~\cite{bowman2}.  For the Laplacian gauge 
fixing, we only use the \lapII gauge~\cite{lapgp}. In this case we construct 
the gauge transformation by 
projecting $M(x)$ constructed from the three lowest lying eigenmodes, 
onto SU(3) by means of trace maximisation. Effectively, 
we maximise the trace of $G(x) M(x) ^{\dagger}$ by iteration over 
Cabibbo-Marinari SU(2) subgroups.

Our numerical calculation begins with an evaluation of the
inverse of $D(\eta)$ with the unfixed gauge configurations, where $D(\eta)$ is 
defined in Eq.~(\ref{D_mu_eqn}).
We approximate the matrix sign function $\epsilon(H_w)$ by the 14th order Zolotarev
approximation~\cite{zolo}. 
We then calculate Eq.~(\ref{overlap_propagator}) for each
configuration and rotate it to Landau or Laplacian gauge by using the 
corresponding gauge transformation
matrices \{$G_i(x)$\}.  Afterward we take the ensemble average to obtain
$S^{\rm bare}(x,y)$.  The discrete Fourier
transformation is then applied to $S^{\rm bare}(x,y)$ and the momentum-space 
bare quark propagator, $S^{\rm bare}(p)$ is obtained finally. 

We use the mean-field improved Wilson action in the overlap fermion kernel. The 
value $\kappa=0.19163$ is used in the 
Wilson action, which provides $m_w a= 1.391$ for the 
regulator mass in the interacting case~\cite{overlgp}. 
We calculate the overlap quark propagator for ten quark
masses on each ensemble by using a shifted Conjugate Gradient solver. For the 
two lattices considered here, the  
quark mass parameter $\eta$ was adjusted to make the tree level bare quark mass
in physical units, the same on the two lattices.  For example, we choose
$\mu = 0.018,$ 0.021, 0.024, 0.030, 0.036, 0.045, 0.060, 0.075, 0.090, and
0.105 on ensemble 1, {\it i.e.}, the $16^3\times{32}$ lattice with $a$ = 0.093 
fm.
This corresponds to bare masses in physical units of
$m^0 = 2 \mu m_w = $ $106$,
$124$, $142$, $177$, $212$, $266$,
$354$, $442$, $531$, and $620$~MeV respectively.


\begin{figure}[h]
\centering\includegraphics[height=0.99\hsize,angle=90]{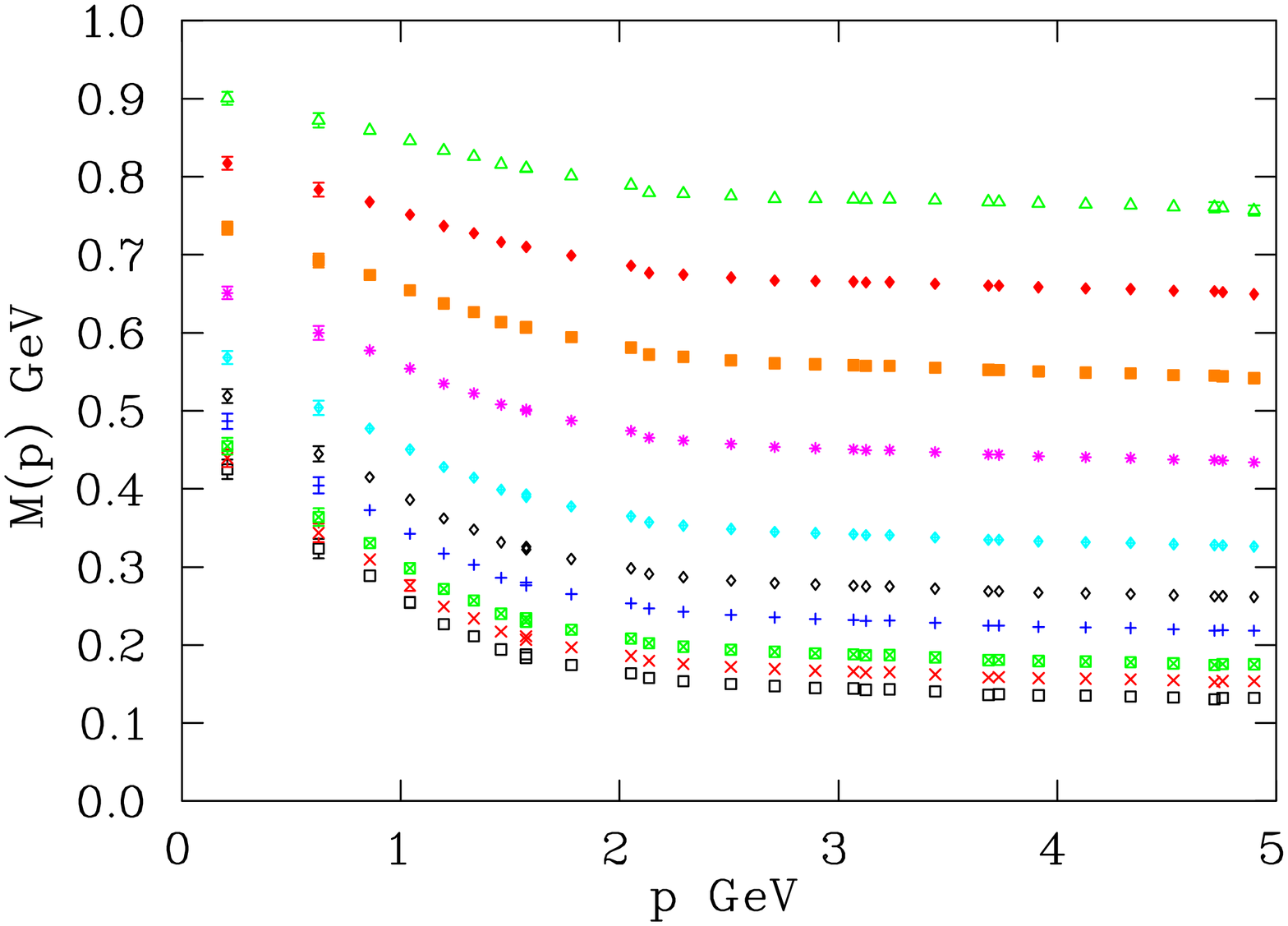}
\centering\includegraphics[height=0.99\hsize,angle=90]{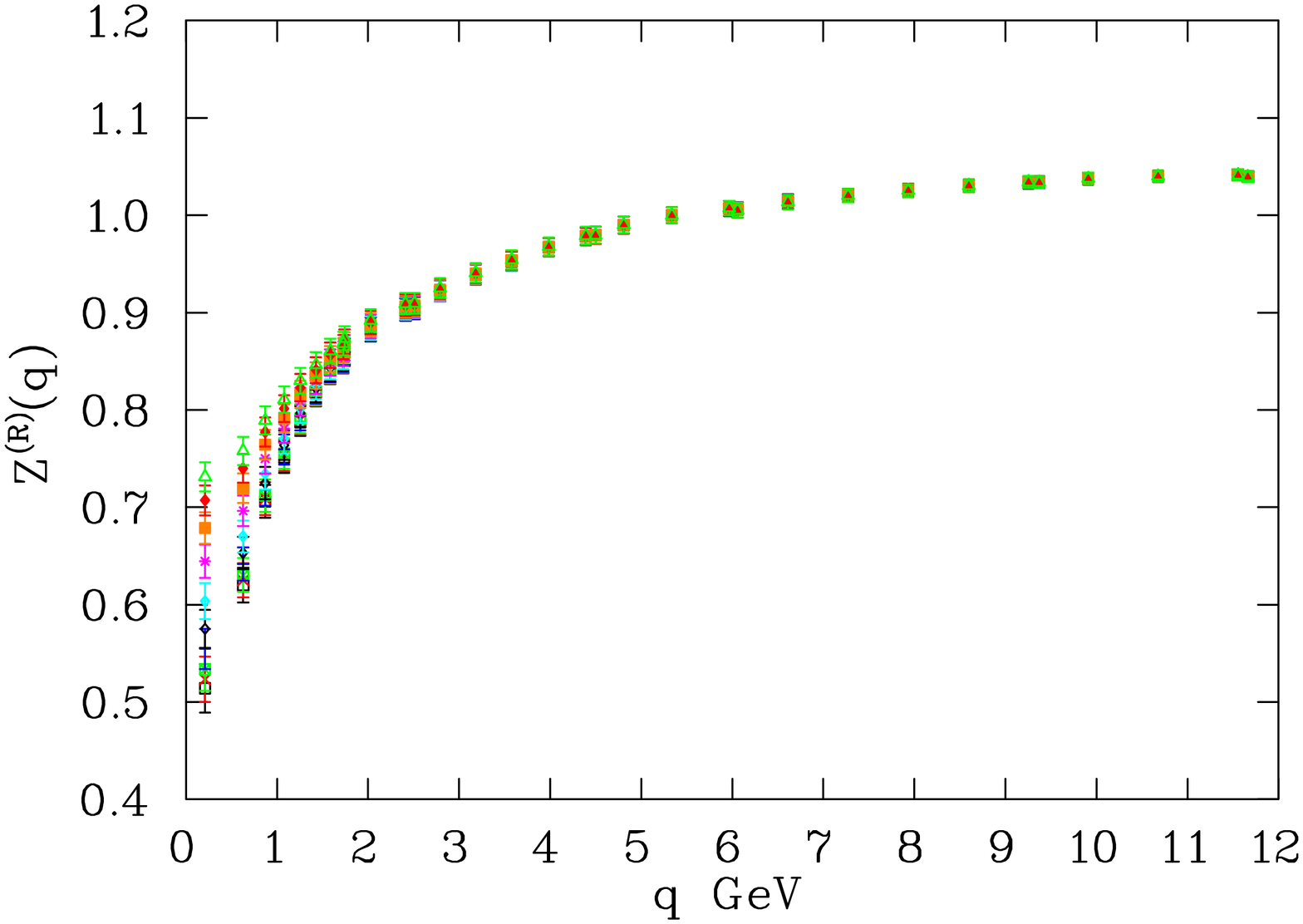}
\caption{(Color online). The functions $M(p)$ and $Z^{(\rm{R})}(q)\equiv Z(\zeta^2;q)$
renormalized at $\zeta=5.31$~GeV (in the $q$ scale ) for all ten quark masses 
in Laplacian gauge on the 
$16^3\times32$ lattice. The mass function $M(p)$ is  plotted
versus the discrete momentum $p$ defined in Eq.~(\ref{dismomt}),
$p=\sqrt{\sum{p_\mu^{2}}}$, over the interval [0,5] GeV,
and $Z^{(\rm{R})}(q)$ is plotted against the kinematic  momentum $q$ defined in
Eq.~(\ref{latmomt}), $q=\sqrt{\sum{q_\mu^{2}}}$, over the interval [0,12] GeV.
The data correspond to bare quark masses (from bottom to top)
$\mu = 0.018,$ 0.021, 0.024, 0.030, 0.036, 0.045, 0.060, 0.075, 0.090, and
0.105, which in physical units correspond to $m^0 = 2\eta m_w \simeq $ 
$106$,
$124$, $142$, $177$, $212$, $266$,
$354$, $442$, $531$, and $620$~MeV respectively
\label{combmovrpq}}
\end{figure}

The results of lattice 2 ($12^3\times 24$) in Landau gauge  were presented in detail in Ref.~\cite{overlgp}, and the
results of lattice 1 ($16^3\times 32$) in Landau gauge were also reported in Ref.~\cite{qpscaling}. 
Here we will focus on the comparison of the results of two lattice gauge fixing
schemes, i.e., the Landau gauge and the Gribov copy free Laplacian gauge, 
to probe the behavior of the overlap fermion propagator with different gauge 
fixings, and the effect of Gribov copies.
Before we make the comparison,  
we first briefly present some data in Laplacian gauge on the $16^3\times32$ lattice, our fine lattice.
All data has been cylinder cut~\cite{Lei99}.  Statistical uncertainties are
estimated via a second-order, single-elimination jackknife.

\subsection{Laplacian gauge}
In Fig.~\ref{combmovrpq} we show the results for all ten
masses for both the mass and wave-function renormalization functions,
$M(p)$ and $Z^{(\rm{R})}(q)\equiv Z(\zeta;q)$ respectively.  As was shown in Ref.~\cite{qpscaling},
the continuum limit is more rapidly approached when the mass function is
plotted against
the discrete lattice momentum $p$, while the wave-function renormalization function
$Z^{(\rm{R})}(q)$ is plotted against the kinematic  momentum $q$.
The renormalization point in Fig.~\ref{combmovrpq} for $Z^{(\rm R)}(q)$ has been
chosen to be $\zeta=5.31$~GeV in the
$q$-scale.

In the plots of $M(p)$, the data is ordered as one
would expect by the values for bare quark mass $m^0$, i.e., the larger the bare quark mass
$m^0$, the higher the $M(p)$ curve.
For large momenta, the function $Z^{(\rm R)}(q)$ demonstrates little mass dependence.
Deviation of $Z^{(\rm R)}(q)$ from its asymptotic value of 1 is a sign of dynamical symmetry
breaking, so we expect the infrared suppression to vanish in the limit of
an infinitely heavy quark.
In the figure for $Z^{(\rm R)}(q)$, the smaller the bare mass, the more pronounced
is the dip at low momenta.
Similarly, at small bare masses $M(q)$ falls off
more rapidly with increasing momenta, which is understood from the fact that a
larger proportion of the infrared mass is due to dynamical chiral
symmetry breaking at small bare quark masses.
These results are much the same as in Ref.~\cite{qpscaling}, which display
the data on the same lattices in Landau gauge.
This qualitative behavior is also consistent with what is seen
in Dyson-Schwinger based QCD models~\cite{agw94,Alkofer}.

\begin{figure}[h]
\centering\includegraphics[height=0.99\hsize,angle=90]{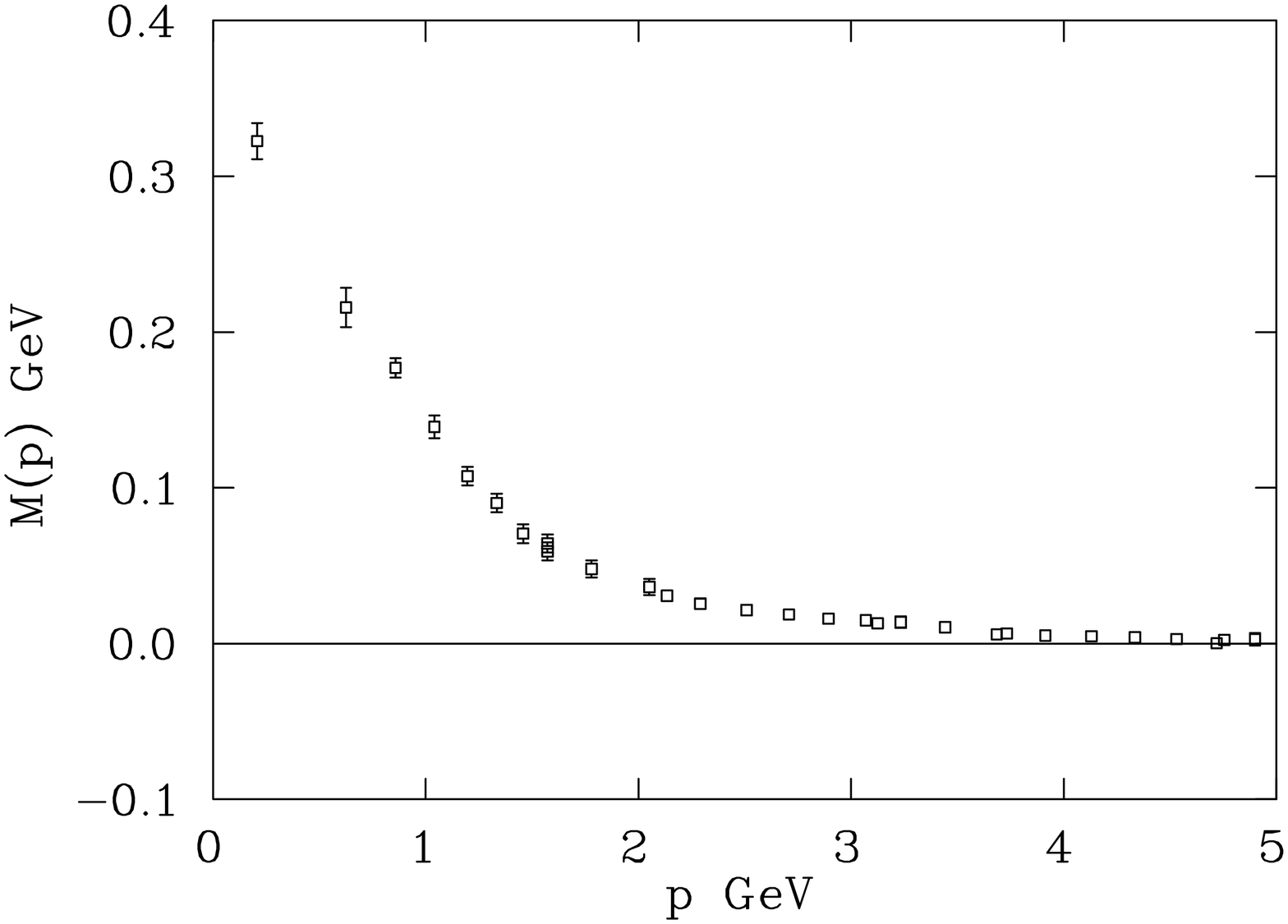}
\centering\includegraphics[height=0.99\hsize,angle=90]{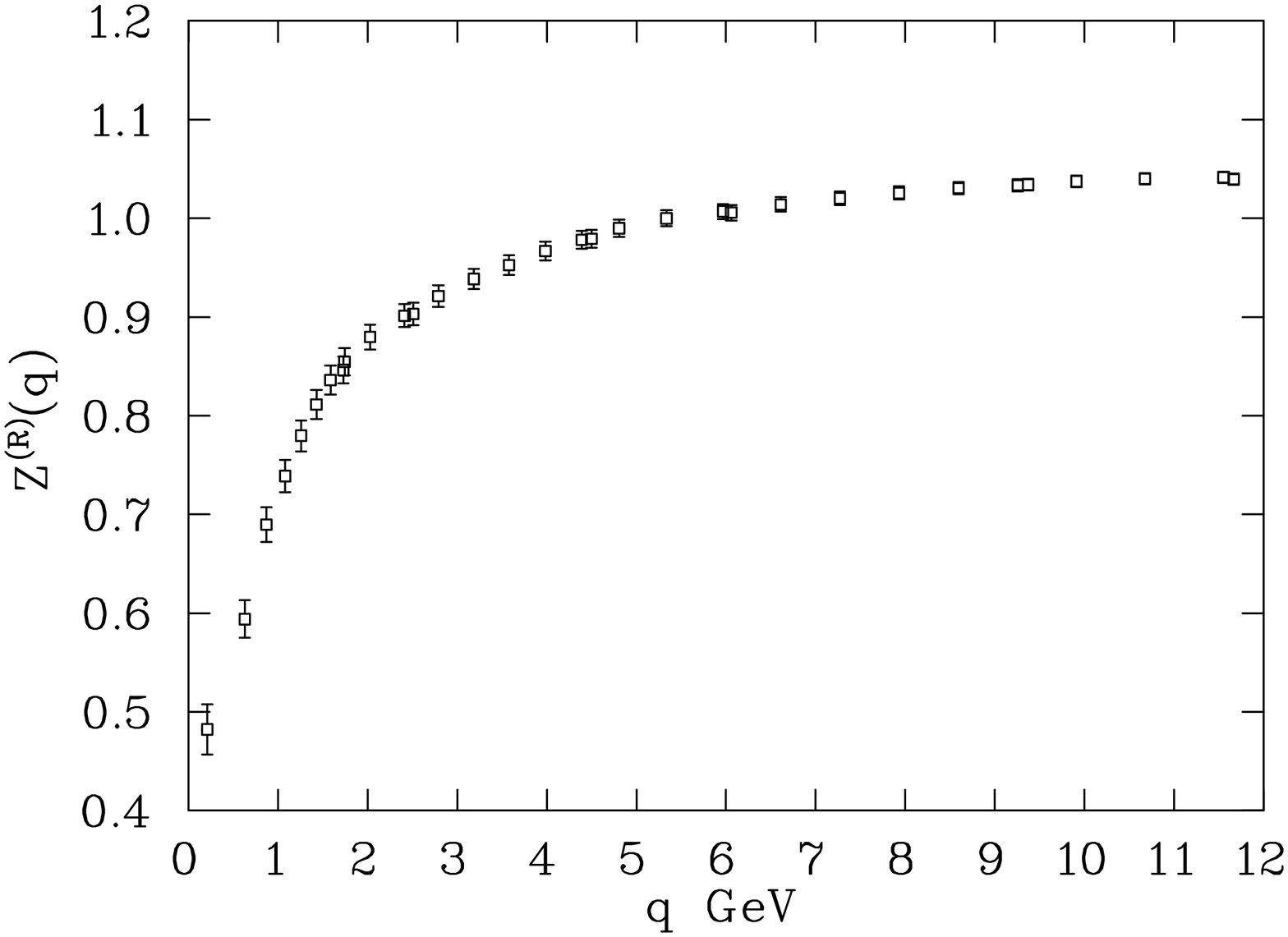}
\caption{The plot of the  functions $M(p)$ and
$Z^{(\rm{R})}(q)\equiv Z(\zeta;q)$ after a linear extrapolation to the chiral limit. 
The mass functions $M(p)$ is plotted against the discrete momentum $p$ in the upper 
part of the figure and  $Z^{(\rm{R})}(q)$ with the renormalization point
$\zeta $ =5.31~GeV (in the $q$-scale) is plotted against the
kinematic momentum $q$ in the lower part of the figure.
\label{extraMZ2pi}}
\end{figure}

In Fig.~\ref{extraMZ2pi} we plot the data
after a linear chiral extrapolation for both functions $M(p)$ and
$Z^{(\rm R)}(q)\equiv Z(\zeta;q)$ in Laplacian gauge.  The mass function $M(p)$ is shown against 
$p$ and while the wave function renormalization function $Z^{(\rm R)}(q)$ is
shown against $q$ with the renormalization point chosen as at 
5.31~GeV in the $q$ scale.
We see that both $M(p)$ and $Z^{(\rm R)}(q)$
deviate strongly from the tree-level behavior, which are $M(p)$ = $m^0$ and $Z^{(\rm R)}(q)$ = 1.  In particular,
as in earlier studies of the Landau gauge quark
propagator\cite{jon1,jon2,Bow02a,overlgp,qpscaling}, we find
a clear signal of dynamical mass generation and
a significant infrared suppression of the $Z(\zeta;q)$ function. 


\begin{figure}[t]
\centering\includegraphics[height=0.99\hsize,angle=90]{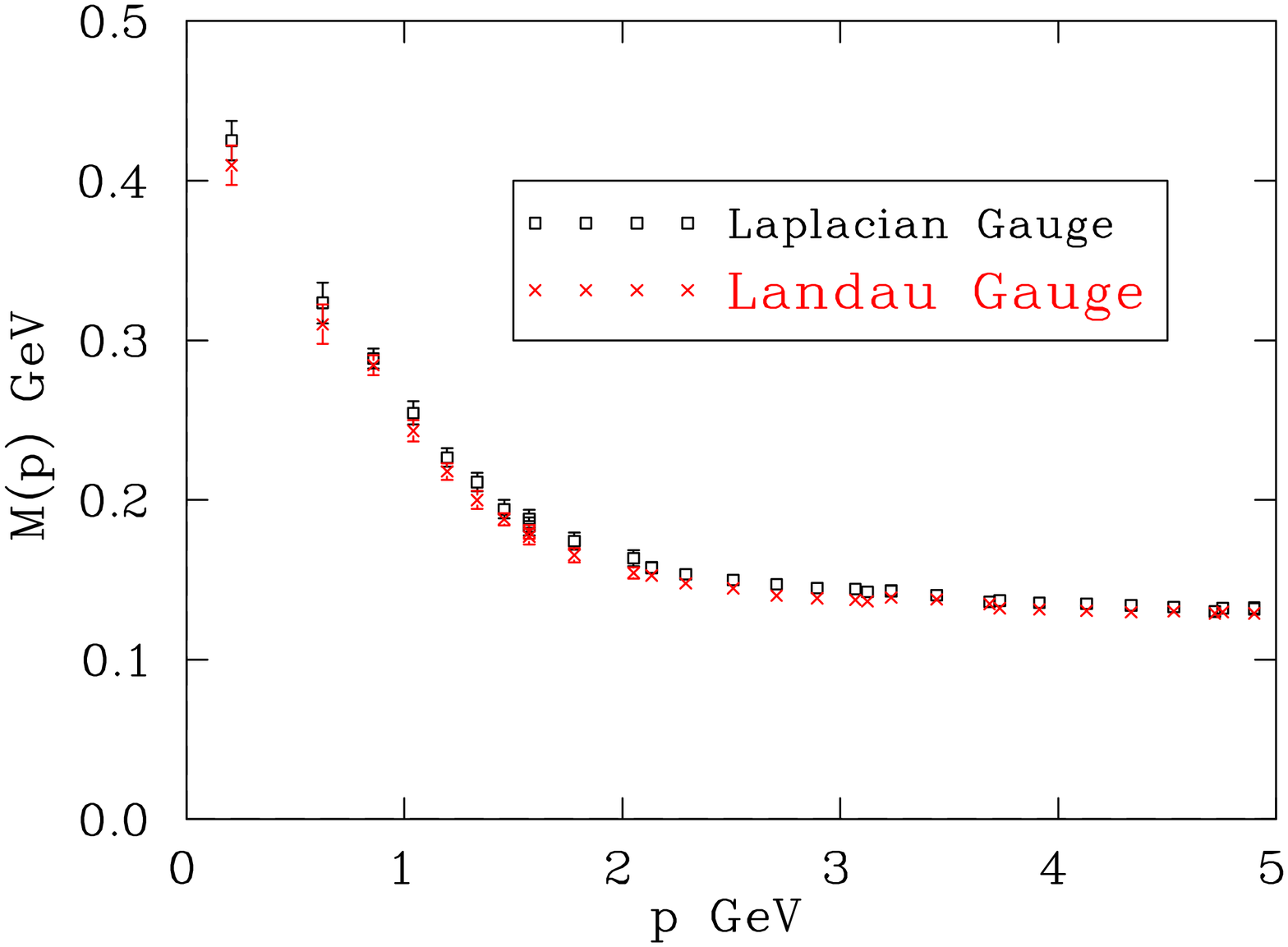}
\centering\includegraphics[height=0.99\hsize,angle=90]{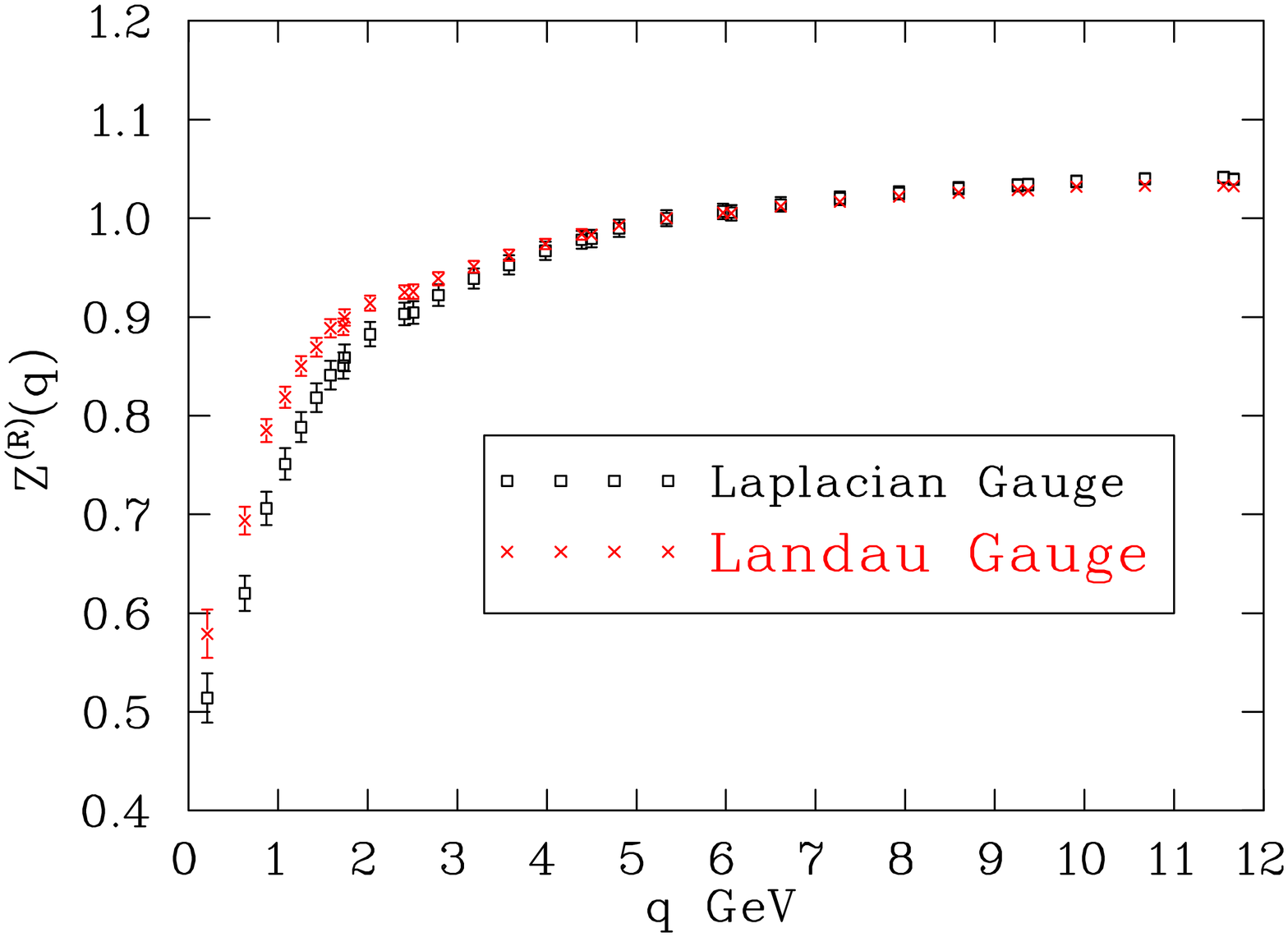}
\caption{(Color online). The comparison of two gauge fixing results at
finite bare quark mass ($m^0$ = 106 MeV) on the fine $16^3\times 32$
lattice with $a$ = 0.093 fm.  $Z^{(\rm{R})}(q)$ is renormalized
to one at the renormalization point $\zeta $ =5.31~GeV (in the
$q$-scale).  For the mass function $M(p)$, Landau gauge and Laplacian
gauge are very similar, while the wave function renormalization
functions $Z^{(\rm{R})}(q)$ differ in the infrared region.  }
\label{compmzsm2g}
\end{figure}

\begin{figure}[h]
\centering\includegraphics[height=0.99\hsize,angle=90]{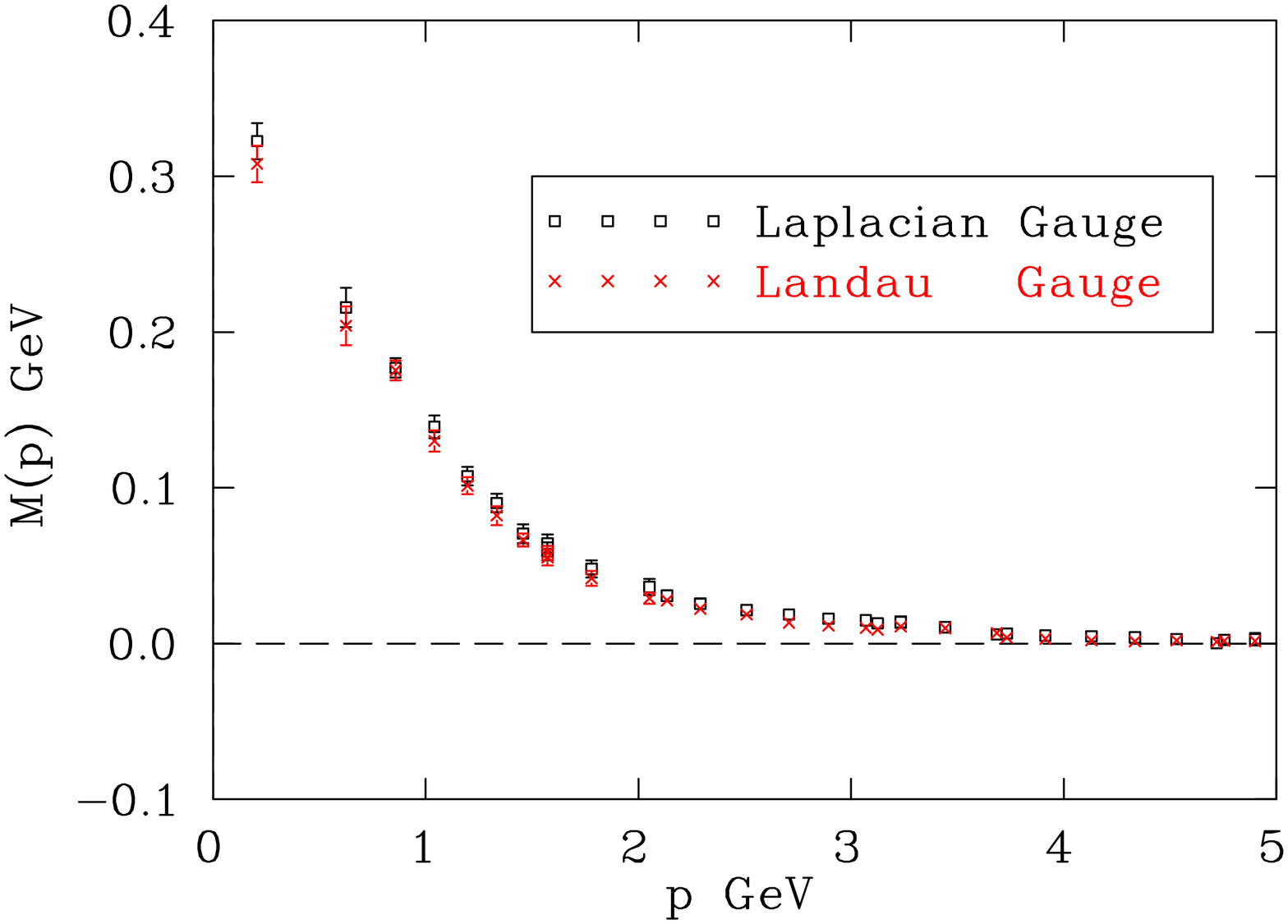}
\centering\includegraphics[height=0.99\hsize,angle=90]{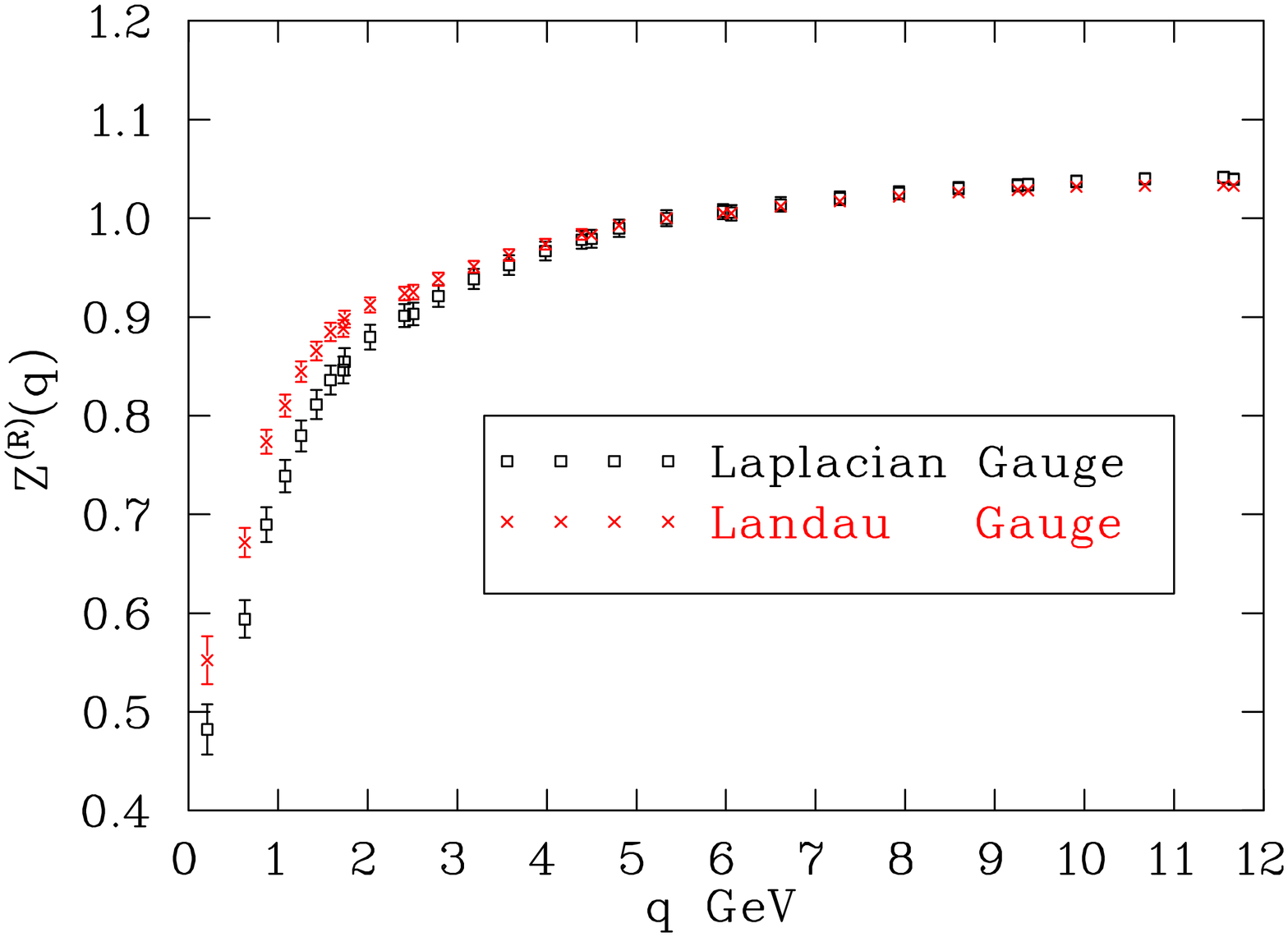}
\caption{(Color online). The comparison of two gauge fixing results after a linear extrapolation to the chiral limit on the fine lattice,
i.e., $16^3\times 32$ at $a$ = 0.093 fm. 
$Z^{(\rm{R})}(q)$ is renormalized to one at the renormalization point
$\zeta $ =5.31~GeV (in the $q$-scale).
For the mass function $M(p)$,  Landau gauge and Laplacian gauge are very similar, 
while the wave function renormalization functions $Z^{(\rm{R})}(q)$ are
similar in the large momentum region  but
differ in the infrared region.
}
\label{compmz12g}
\end{figure}

\begin{figure}[h]
\centering\includegraphics[height=0.99\hsize,angle=90]{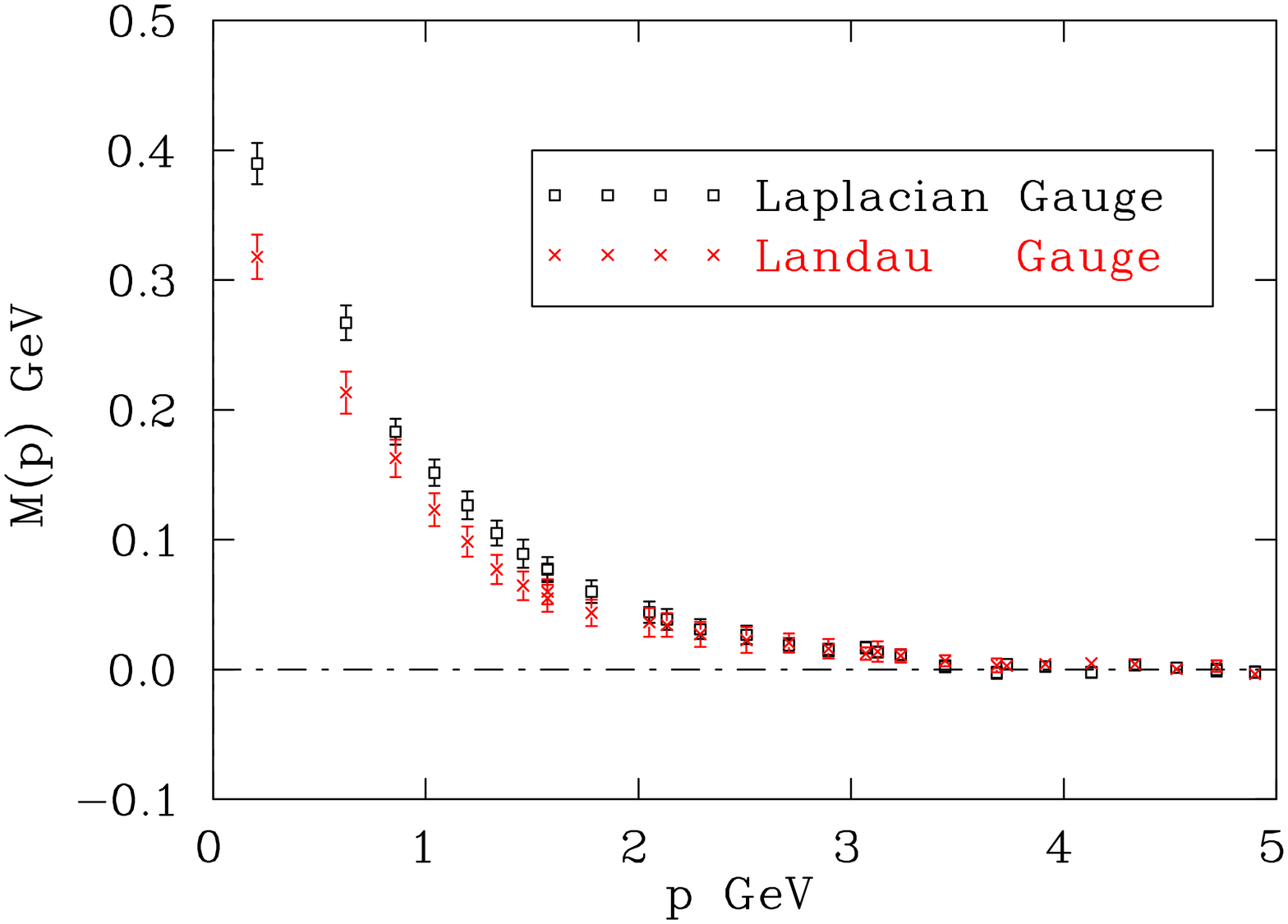}
\centering\includegraphics[height=0.99\hsize,angle=90]{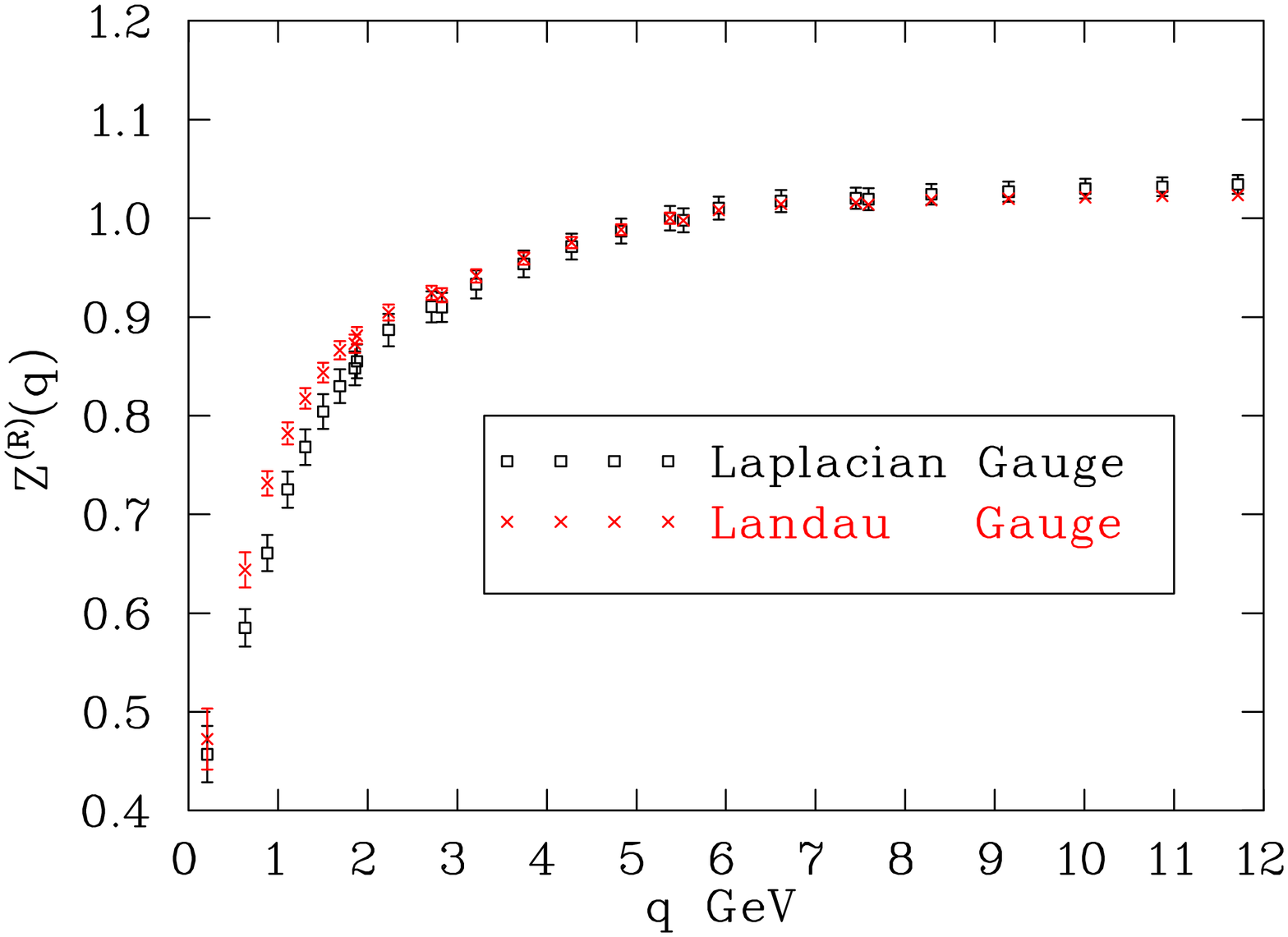}
\caption{(Color online). The comparison of two gauge fixing results in the chiral limit on the coarse lattice, 
i.e., $12^3\times 24$ with $a$ = 0.124 fm. 
The small gauge dependence of the infrared behavior of the Z-function is similar to that on
the fine lattice in Fig.~\ref{compmz12g}. The infrared mass functions, $M(p)$, appear different on this 
coarse lattice whereas they were similar on the fine lattice. This suggests larger \oa{2} errors in 
Laplacian gauge.
\label{compmz22g}}
\end{figure}
\subsection{Gauge fixing comparison}

Next we present the results in the two gauge fixing schemes for
comparison.  All data have been cylinder cut~\cite{Lei99}.  First we
give the results on the $16^3\times 32$ lattice.

Fig.~\ref{compmzsm2g} reports results for the mass and renormalization
functions at our lightest bare quark mass of $m^0 = 106$ MeV.  These
results may be compared with the Asqtad results of Ref.~\cite{Bow02a},
where Fig.~9 compares results of various gauge fixing schemes for the
renormalization function.  There, Landau gauge results are seen to lie
significantly higher than the Laplacian gauge results in the infrared.
However, with overlap fermions, Fig.~\ref{compmzsm2g} indicates the
Landau gauge results lie much closer to the Laplacian gauge results.
Given the improved chiral properties of the overlap operator, the
present results should provide a better indication of the continuum
limit behavior.  In either case, the same qualitative behavior of
Landau Gauge sitting above Laplacian gauge in the infrared is
observed.

The mass function of Fig.~\ref{compmzsm2g} reveals an approximate
invariance on the selection of Landau or Laplacian gauges.  Figs.~12
and 13 of Ref.~\cite{Bow02a} indicate that the mass function of
Asqtad fermions is also insensitive to the choice of Landau gauge or
the Gribov-copy free Laplacian gauge.  

We now proceed to compare the data in the chiral limit.
Fig.~\ref{compmz12g} shows the comparison of the mass function,
$M(p)$, and the wave function renormalization function,
$Z^{(\rm{R})}(q)$, in Landau gauge and Laplacian gauge.  We see that
they give similar performance in terms of rotational symmetry and
statistical noise.  Looking more closely, we can see that Landau gauge
gives a slightly cleaner signal at this lattice spacing.  We also note
that at very large momenta, the two gauge fixing schemes give similar
results as expected.  Although Laplacian gauge is a non-local gauge
fixing scheme and difficult to understand perturbatively, it is
equivalent to Landau gauge in the asymptotic region~\cite{baal95}. In
the infrared region, the mass function, $M(p)$, in the two gauges are
very similar.  For the mass functions there is a hint that the data in
Laplacian gauge are a little higher than for the Landau gauge,
although they agree within statistical errors. With greater statistics
we may resolve a small difference.  For the renormalization function
$Z^{(\rm{R})}(q)$, there are systematic differences in the infrared
region. The $Z^{(\rm{R})}(q)$ is more strongly infrared suppressed in
the Laplacian gauge than in the Landau gauge.  That is consistent with
what was seen in the case of the Asqtad quark action~\cite{Bow02a}
when comparing these two gauges.

Now we give the results from the coarse lattice.   
Fig.~\ref{compmz22g} shows the comparison of the mass function $M(p)$ and the wave function
renormalization function $Z^{(\rm{R})}(q)$ in the chiral limit in Landau gauge and
Laplacian gauge on the $12^3\times 24$ lattice with $a$ = 0.124 fm.
As in the case of the $16^3\times 32$ lattice, at very large momenta, the two gauge fixing schemes give similar results.
For the renormalization function $Z^{(\rm{R})}(q)$, the situation is very similar to the case
of  the $16^3\times 32$ lattice, i.e., at very large momenta, the two gauge fixing schemes give similar results,
while in the infrared region, $Z^{(\rm{R})}(q)$ is more strongly 
suppressed in Laplacian gauge than in Landau gauge.
For the mass function, $M(p)$, the situation is different to that seen on the fine lattice. 
In the infrared region, the data in Laplacian gauge sit higher than in Landau gauge.
While this is consistent with the infrared behavior of the renormalization function, $Z^{(\rm{R})}(q)$,
which is more suppressed in Laplacian gauge than in  Landau gauge, there is no similar signal
for our fine lattice $16^3\times 32$. In that case (see Fig.~\ref{compmz22g}), the data in Laplacian gauge agree with that in
Landau gauge within error bars, although there is a hint that the data in Laplacian gauge are
a little higher than in Landau gauge in the infrared region. This infrared behavior of the mass function in Laplacian gauge
is likely caused by the finite lattice spacing errors, i.e., on fine enough lattices we expect Landau and Laplacian gauge results for
 $M(p)$ to be very similar.


\begin{figure}[t]
\centering\includegraphics[height=0.99\hsize,angle=90]{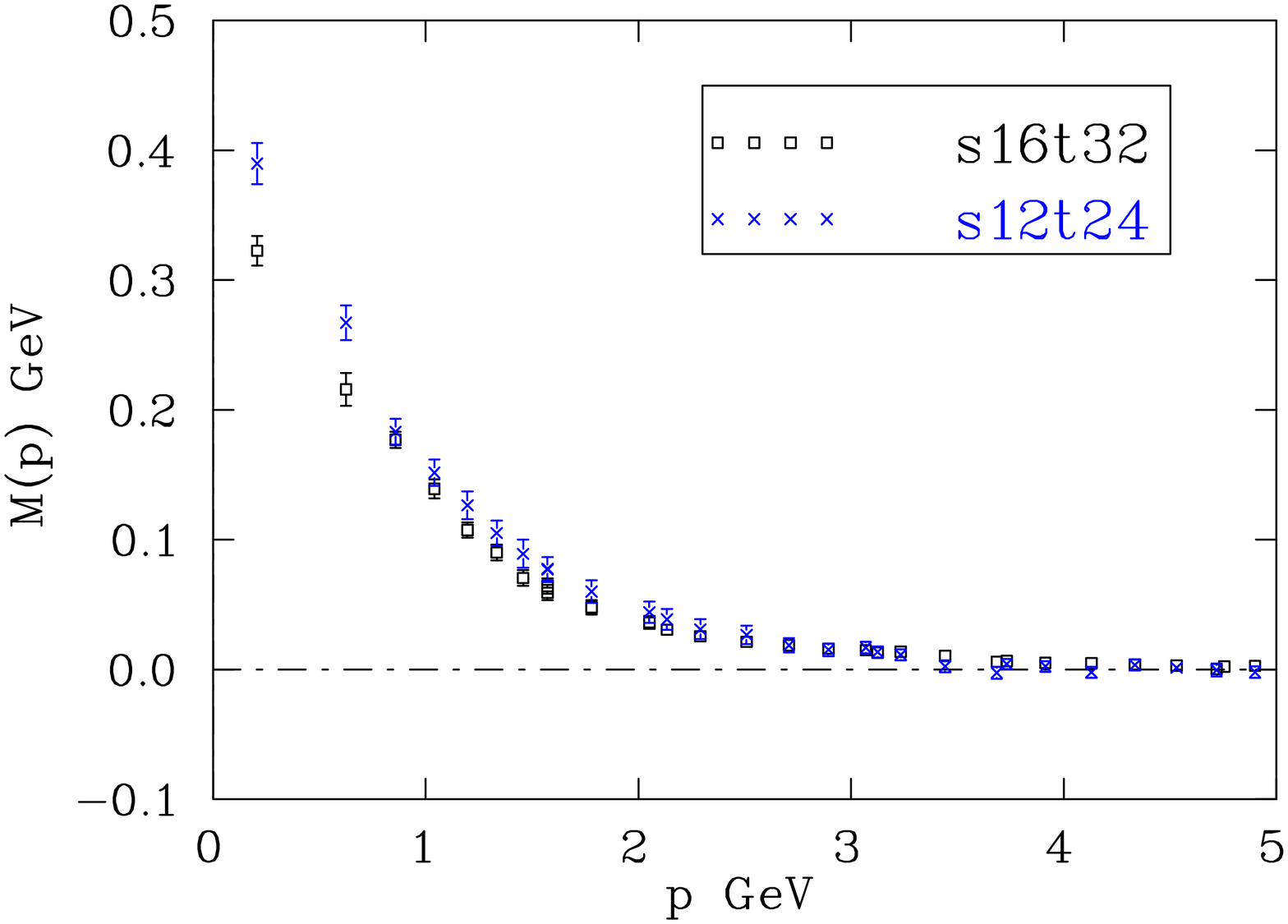}
\centering\includegraphics[height=0.99\hsize,angle=90]{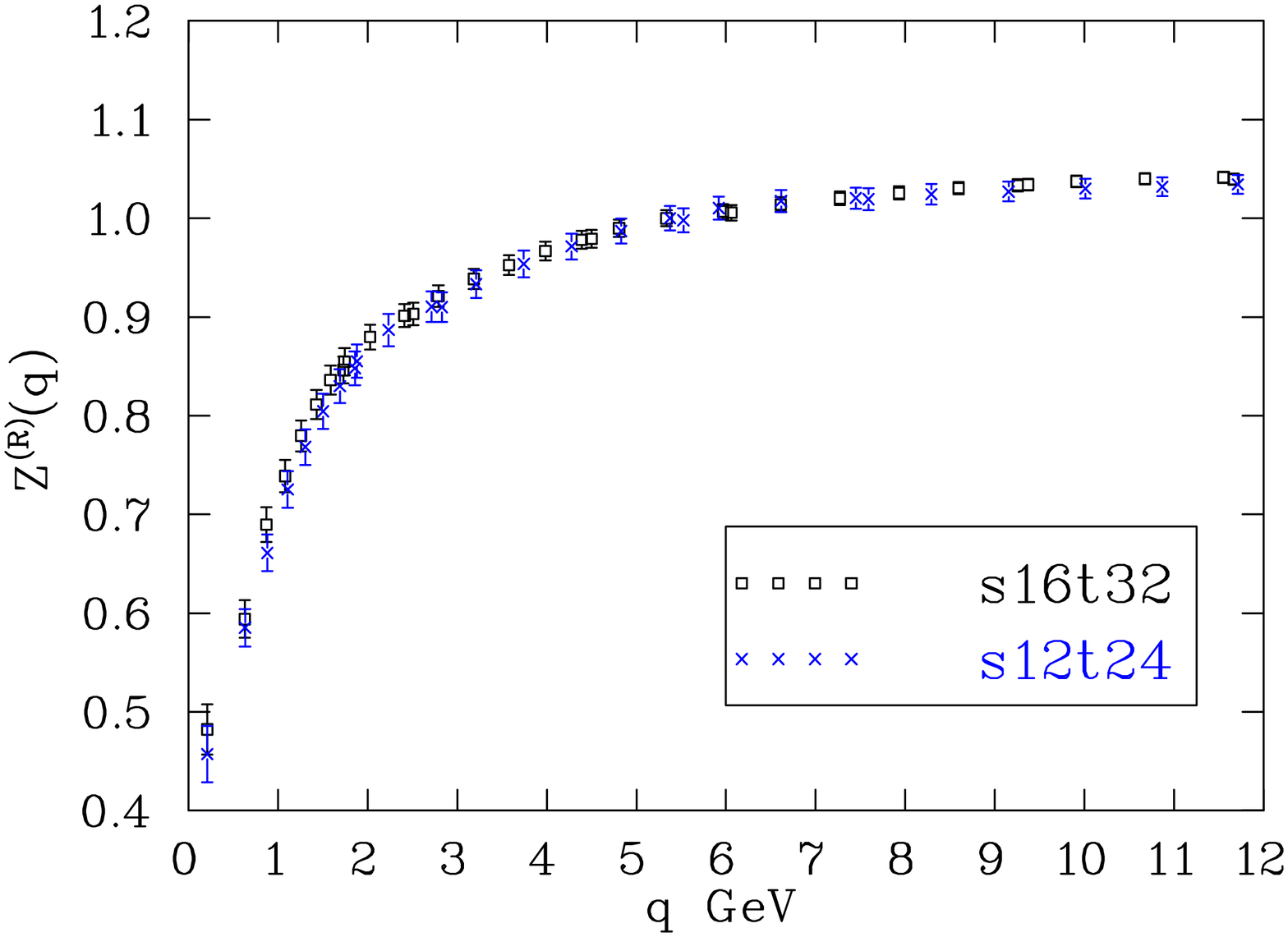}
\caption{(Color online). The comparison of Laplacian gauge results in the chiral limit on two lattices,
i.e., $12^3\times 24$ with $a$ = 0.124 fm and $16^3\times 32$ with $a$ = 0.093 fm.
The renormalization point for $Z^{(\rm{R})}(q)\equiv Z(\zeta;q)$
is chosen to be $\zeta $ =5.31~GeV (in the $q$-scale).
For the mass functions $M(p)$, the value on $12^3\times 24$  lattice is higher
than that on the $16^3\times 32$ lattice in the infrared region. 
while the wave function renormalization function $Z^{(\rm{R})}(q)$ on two lattices
agree within error bars. This suggests that the finite $a$ errors are not small on the
coarse lattice in Laplacian gauge. 
\label{comp2lp}}
\end{figure}
Finally, we  present the data for the two lattices in Laplacian gauge to further explore 
possible finite $a$ errors.                                                
Fig.~\ref{comp2lp} shows the comparison of the  mass function, $M(p)$, and the wave function
renormalization function $Z^{(\rm{R})}(q)$ in the chiral limit in 
Laplacian gauge on the $12^3\times 24$ and the $16^3\times 32$ lattices.
For the renormalization function, $Z^{(\rm{R})}(q)$,
results from the two lattices have small differences, but agree with each other 
within errors.
For the mass function, $M(p)$, the results agree
well at large momenta, but there is a substantial difference in the infrared region.
A similar comparison was made between these  two lattices in Landau gauge~\cite{qpscaling}.
In that case, both the renormalization function, $Z^{(\rm{R})}(q)$, and the mass function, $M(p)$, 
agree well on the two lattices. This indicates that in Landau gauge, the finite $a$ errors are small
even for our coarse lattice with lattice spacing $a$ = 0.124 fm.  But in Laplacian gauge, 
the  finite $a$ errors are not negligible for our coarse lattice  $12^3\times 24$.  
A likely explanation for this is the fact that our implementation of the Laplacian gauge
is not \oa{2} improved.

\section{summary and outlook}

The momentum-space quark propagator has been studied in Landau gauge as well
as the Gribov copy free Laplacian gauge on two lattices with the same physical volume but
with different lattice spacings $a$ .
We calculated the nonperturbative momentum-dependent wave-function
renormalization, $Z(q)$, and the nonperturbative mass function, $M(p)$, for a
variety of bare quark masses.  We also performed a simple linear extrapolation
to the chiral limit.

At very large momenta the two gauge-fixing schemes give similar results as
expected.  Laplacian gauge is equivalent to the Landau gauge in the
asymptotic region.  In the infrared region, the mass function, $M(p)$,
in the two gauges are very similar on the fine lattice, but differ on the 
coarse lattice.  The present Laplacian gauge fixing is not \oa{2} improved. 
For our fine lattice in the infrared region, the mass
function, $M(p)$, agrees within statistical errors in
the two gauge fixings. However, there is a hint that  the data in Laplacian
gauge may be a little higher than in Landau gauge, which would be consistent 
with the behavior of the renormalization function, $Z^{(\rm{R})}(q)$.
For the renormalization function, $Z^{(\rm{R})}(q)$, there are systematic 
differences in the infrared region.
The renormalization function is more strongly infrared suppressed in the 
Laplacian gauge than in the Landau gauge.

\section{ACKNOWLEDGMENTS}
We thank the Australian National Computing Facility for Lattice Gauge
Theory and both the Australian Partnership for Advanced Computing (APAC)
and the South Australian Partnership for Advanced Computing (SAPAC)
for generous grants of supercomputer time which have enabled this
project.  This work is supported by the Australian Research Council.

\end{document}